\documentstyle[aps,preprint,eqsecnum,floats,epsf]{revtex}
\tighten
\draft
\begin{document}

\preprint{}

\title{Character Expansions for the 
Orthogonal and Symplectic Groups}

\author{A.B. Balantekin\thanks
{Electronic address: {\tt baha@nucth.physics.wisc.edu}}}
\address{Department of Physics, University of Wisconsin\\
         Madison, Wisconsin 53706 USA\thanks{Permanent Address},\\
and\\
Max-Planck-Institut f\"ur Kernphysik,
Postfach 103980, D-69029 Heidelberg, Germany}
\author{P. Cassak\thanks
{Electronic address: {\tt pcassak@nucth.physics.wisc.edu}}}

\address{Department of Physics, University of Wisconsin\\
         Madison, Wisconsin 53706 USA}


\maketitle

\begin{abstract}

Formulas for the expansion of arbitrary invariant group functions in
terms of  the characters for the $Sp(2N)$, $SO(2N+1)$, and $SO(2N)$
groups are derived using a combinatorial method.  The method is
similar to one used by Balantekin to expand group functions over the
characters of the $U(N)$ group.  All three expansions have been
checked for all $N$ by using them to  calculate the known expansions
of the generating function of the homogeneous  symmetric functions. An
expansion of the exponential of the traces of  group elements,
appearing in the finite-volume gauge field partition functions, is 
worked out for the orthogonal and symplectic groups. 

\end{abstract}


\vglue1cm

\section{Introduction}

The expansion of invariant functions of a group into its characters
(traces of the representation matrices) \cite{Balantekin:2000vn} is
very useful in a number of physical situations. In $U(N)$ lattice
gauge theories and in the lattice expansion of the non-linear $U(N)
\times U(N)$ sigma model calculation of certain $U(N)$ group integrals
are needed
\cite{Bars:1979xb,Bars:1980yy,Samuel:1980vk,Brezin:1980rk,%
Brower:1981vt}.
If the integrands can be expanded in terms of the $U(N)$ characters
then such integrals can easily be calculated
\cite{Bars:1980yy,Bars:1981bn,Balantekin:1984km}. Similar $U(N)$
integrals also arise in the statistical theory of nuclear reactions
\cite{mello}. In 1980 Itzykson and Zuber calculated a particular
unitary group integral \cite{Itzykson:1980fi} which turned out to be a
special case of a more general formula by Harish-Chandra
\cite{chandra}. The Itzykson-Zuber integral and its generalizations
\cite{berezin,Guhr:1996vx,Jackson:1996jb,Smilga:1995tb} are also
easily dealt with using character expansions
\cite{Balantekin:2000vn}. 

The character  expansion of an invariant function of group elements 
is given by 
\begin{equation}
f (\det U, Tr U, \cdots) = \sum_r a_r \chi_r (U), 
\end{equation}
where $\chi_r (U)$ is the character of the representation $r$.  
Since group characters form an orthogonal set \cite{murn} the
coefficients can be
obtained by explicitly integrating the product of this function with
the characters over the group manifold: 
\begin{equation}
a_r = \int dU \chi_r^* (U) f (\det U, Tr U, \cdots).
\end{equation}
(Note that $a_0$ is the integral of the function itself over the 
group manifold). It is rather difficult
to obtain complicated character expansions by explicit integration. 
In 1984 Balantekin \cite{Balantekin:1984km} developed a combinatorial
method that enabled one to solve for the coefficients in some
expansions over the $U(N)$ group characters that was quite simple in
comparison to performing group integrals. Needing a more general
version, the result was recently extended in its range of
applicability \cite{Balantekin:2000vn}. 

In a parallel development it was shown  that the spectral density of 
the Dirac operator for a gauge theory near its zero eigenvalues 
should only depend on the symmetries in question 
\cite{Shuryak:1993pi,Verbaarschot:1993pm,Verbaarschot:1994qf}.
Although the original work
\cite{Shuryak:1993pi,Verbaarschot:1993pm,Verbaarschot:1994qf} used a 
Gaussian random matrix model, the results from the random matrix 
theory can be proven to be universal 
\cite{Akemann:1997vr,Sener:1998vj,Klein:2000pj,Damgaard:1998ye,%
Akemann:2000ze}. 
This implies that the spectral density of the Dirac operator
near the origin can be extracted from random matrix theories which
provide a description of common aspects of various quantum phenomena
(for a review see \cite{Guhr:1998ve}). Hence to study the low-energy
limit of, for example, quantum chromodynamics (QCD) one needs to 
choose a random matrix theory with the global symmetries of the QCD
partition function. The partition functions  
calculated from the effective field theory and random matrix theory
agree \cite{Shuryak:1993pi,Halasz:1995qb,Damgaard:1998nd}. 
These random matrix theories are characterized by
the Dyson index $\beta$ which is the number of independent variables
per matrix element \cite{Verbaarschot:1994qf,dyson}. For fermions in 
the fundamental 
representation $\beta =1$ for $N_c=2$ and $\beta = 2$ for $N_c \ge 2$
where $N_c$ is the number of colors. For fermions in the adjoint
representation and $N_c \ge 2$ we have $\beta =4$. For $\beta=2$ the
low-energy (finite-volume) QCD partition function is the same as the
one-link integral of two-dimensional lattice QCD
\cite{Brezin:1980rk,Gross:1980he} and is calculated using the $U(N)$
character expansion \cite{Balantekin:2000vn} and other methods
\cite{Leutwyler:1992yt,Jackson:1996jb,Akuzawa:1998tq}. For $\beta =4$
the zero momentum Goldstone modes belong to the coset space
$SU(N_f)/SO(N_f)$ \cite{Leutwyler:1992yt,Smilga:1995tb} where $N_f$ is
the number of fermion flavors. Hence the finite-volume partition
function is a group integral where the argument is the exponential of
an $SO(N_f)$ group element. Similarly in the $\beta=1$ case the coset
space of the Goldstone modes is $SU(2N_f)/Sp(2N_f)$ 
\cite{Leutwyler:1992yt,Smilga:1995tb}. Massive partition functions of 
random matrix ensembles 
with $\beta=1$ and $4$ were considered in Refs. \cite{Akemann:2000ze}
and \cite{Nagao:2000qn}. To calculate 
these partition functions it is very useful to have expressions for 
character expansions over the orthogonal and symplectic 
groups. Explicit expressions for these partition functions, for 
example,  may help in finding solutions of Virasoro constraints which
were found so far only for the $\beta=2$ case. (For the application of
Virasoro constraints on the effective finite volume partition function
see, for example,
Refs. \cite{Damgaard:2000ic,Akemann:2001df,Dalmazi:2001hd}). 

The present work is an extension of Balantekin's method of finding the
expansion coefficients for expansions over the characters for the
symplectic group $Sp(2N)$, the odd dimensional special orthogonal
group $SO(2N+1)$, and the even dimensional special orthogonal group
$SO(2N)$. Some background material will be treated in Section
\ref{sec-back}, including general information about the characters of
the groups in question.  The procedures for developing the expansions
for the different groups are similar, so Section \ref{sec-derive} will
treat the general idea.  The specific expressions will be derived in
Section \ref{sec-sp2n} for $Sp(2N)$, Section \ref{sec-so2n1} for
$SO(2N+1)$, and Section \ref{sec-so2n} for $SO(2N)$. Finally, some
examples of expansions will be given for each of the groups in Section
\ref{sec-example}.  

For quick reference, the expansions and the most general expressions
for their coefficients for $Sp(2N)$, $SO(2N+1)$, and $SO(2N)$ are
found in Equations (\ref{sp2nexp}) and (\ref{sp2ngen}),
(\ref{son1exp}) and (\ref{so1gen}), and (\ref{so2nexp}) and
(\ref{so2gen}), respectively. 

\section{Background and Formulas for  Characters}
\label{sec-back} 

In order to calculate expressions for the character expansions, we
will need expressions for the characters.  These characters have been
furnished by Weyl \cite{weyl}.  For reference, the Weyl formulas for
the $U(N)$, $Sp(2N)$, $SO(2N+1,\Re)$, and $SO(2N,\Re)$ group
characters are reprinted below.  (We have included the formula for the
$U(N)$ group characters for completeness even though we will not need
them in the present work.)

In the following, $\det[{\cal A}_{ij}]$ refers to the determinant of
the $N \times N$ matrix ${\cal A}$ whose entry in the $i$-th row and
$j$-th column is ${\cal A}_{ij}$.  Furthermore, we will denote a
matrix in the unitary group as $U$, the symplectic group as $P$, and
the orthogonal group as $R$, and the eigenvalues of any of these
matrices are labeled by $t_{i}$.  For the $U(N)$ case, there are $N$
eigenvalues that are all phases.  For the $Sp(2N)$ and $SO(2N)$ cases,
there are $2N$ eigenvalues, but they come in pairs of a phase and its
reciprocal.  Thus, a complete list of eigenvalues would be $t_{1},
t_{2}, \ldots, t_{N}, t_{1}^{-1}, t_{2}^{-1}, \ldots, t_{N}^{-1}$.
For the $SO(2N+1)$ case, it is the same as the even dimensional cases
with an additional eigenvalue of $t=1$.  The determinants given below,
then, are determinants of $N \times N$ matrices which contain
functions of only the individual eigenvalues, not their reciprocals.
Finally, the character is a function of the representation, which is
labeled by a partition  $(n_{1}, n_{2}, \ldots, n_{N})$ where the
non-negative integers $n_{i}$ satisfy $n_{1} \geq n_{2} \geq \ldots
\geq n_{N}$.  Each representation corresponds to a permissible Young
Tableau.

The expressions for the simple characters of the $U(N)$,  $Sp(2N)$,
and $SO(2N+1)$ groups are
\begin{equation}
\chi_{(n_{1}, n_{2}, \ldots, n_{N})}(U) = \frac{\det\left[t_{i}^
{n_{j}+N-j}\right]}{\det\left[t_{i}^{N-j}\right]},
\end{equation}
\begin{equation}
\chi_{(n_{1}, n_{2}, \ldots, n_{N})}(P) = \frac{\det\left[t_{i}^
{n_{j}+N-j+1} - t_{i}^{-(n_{j}+N-j+1)}\right]}
{\det\left[t_{i}^{N+1-j}-t_{i}^{-(N+1-j)}\right]},   \label{sp}
\end{equation}
and
\begin{equation}
\chi_{(n_{1}, n_{2}, \ldots, n_{N})}(R) = \frac{\det\left[t_{i}^
{n_{j}+N-j+\frac{1}{2}} - t_{i}^{-(n_{j}+N-j+\frac{1}{2})}\right]}
{\det\left[t_{i}^{N+\frac{1}{2}-j} -
t_{i}^{-(N+\frac{1}{2}-j)}\right]},       \label{on1}
\end{equation}
respectively.

The $SO(2N)$ case requires more attention.  We define
\begin{equation}
{\cal C}_{(n_{1}, n_{2}, \ldots, n_{N})}(R) = \frac{
\det\left[t_{i}^{n_{j}+N-j} + t_{i}^{-(n_{j}+N-j)} -
\delta_{jN}\delta_{n_{N}0}\right]}{\det\left[t_{i}^{N-j} +
t_{i}^{-(N-j)} - \delta_{jN}\right]},  \label{on2}
\end{equation}
and
\begin{equation}
{\cal S}_{(n_{1}, n_{2}, \ldots, n_{N})}(R) = \frac{
\det\left[t_{i}^{n_{j}+N-j} - t_{i}^{-(n_{j}+N-j)}\right]}
{\det\left[t_{i}^{N-j} + t_{i}^{-(N-j)} - \delta_{jN}\right]}.
\end{equation}
(The notation in the previous two equations is non-standard, but they
give the proper elements as stated in Ref. \cite{murn} in a more
modern  and manipulable form.)   ${\cal C}_{(n_{1}, n_{2},
\ldots,n_{N})}(R)$ alone is the simple character of $SO(2N)$ if and
only if $n_{N}=0$.  If $n_{N} \neq 0$, then ${\cal C}_{(n_{1}, n_{2},
\ldots, n_{N})}(R)$ is a double character.  For this case, the simple
characters are given by $\frac{1}{2}  ({\cal C}_{(n_{1}, n_{2},
\ldots, n_{N})}(R)  \pm {\cal S}_{(n_{1}, n_{2}, \ldots, n_{N})}(R))$.
In the present work, only the expression for ${\cal C}_{(n_{1}, n_{2},
\ldots, n_{N})}(R)$ given in Eq. (\ref{on2}) will be needed.  This
statement will be justified in Section \ref{sec-so2n} where $SO(2N)$
is treated.

One last property of these expressions that will be useful for
checking the reliability of the expansions derived in this paper is
the value of the characters for representations corresponding to Young
tableaux of one row ($n_{1}=n$, all others are 0) and one column
($n_{n}=1$ for all $n$ up to some value, all others are 0).  The
characters for representations with one row, labeled $(n)$, and one
column, labeled $(1^{n})$, are 
\begin{eqnarray}
& \chi_{(n)}(U) = h_{n}(t_{i}),&  \chi_{(1^{n})}(U) = a_{n}(t_{i}), \\
& \chi_{(n)}(P) = h_{n}(t_{i}, t_{i}^{-1}),& \chi_{(1^{n})}(P) =
a_{n}(t_{i},t_{i}^{-1})-a_{n-2}(t_{i},t_{i}^{-1}),   \label{hasp} \\ &
\chi_{(n)}(R) =
h_{n}(t_{i},t_{i}^{-1},1)-h_{n-2}(t_{i},t_{i}^{-1},1),&
\chi_{(1^{n})}(R) = a_{n}(t_{i},t_{i}^{-1},1),    \label{haso2n1} \\ &
{\cal C}_{(n)}(R) =
h_{n}(t_{i},t_{i}^{-1})-h_{n-2}(t_{i},t_{i}^{-1}),&  {\cal
C}_{(1^{n})}(R) = a_{n}(t_{i},t_{i}^{-1}),   \label{haso2n} 
\end{eqnarray}
for the $U(N)$, $Sp(2N)$, $SO(2N+1)$, and $SO(2N)$ groups,
respectively. The functions $h_{n}(t_{i})$ are the homogeneous
symmetric functions of order $n$, and the functions $a_{n}(t_{i})$ are
the elementary symmetric functions of order $n$.  Further discussion
is given in Ref. \cite{Balantekin:2000vn}. 

\section{General Properties of the Derivation}
\label{sec-derive}

We are now ready to derive the form for the expansions of group
functions over the characters of the various groups mentioned in the
previous section.  As with any expansion, the crux of this issue is
being able to determine and calculate the coefficients in the
expansion.  The goal of the next four sections will be to find these
coefficients.  

The derivation is very similar to the one used by Balantekin
\cite{Balantekin:2000vn} in finding the coefficients of the expansion
over the unitary group characters.  As the expressions for the group
characters for $Sp(2N)$, $SO(2N+1)$, and $SO(2N)$ are all similar, the
derivations for  all three will proceed in much the same manner.  To
make the general method more transparent, the common aspects of the
derivation will be presented in this section without mention of the
specific groups.  The following three sections will be devoted to
using the result of this section to derive expressions for the
coefficients in the expansions over the characters in each group. 

We begin by noting that each of the expressions for the characters
given by Eqs. (\ref{sp}), (\ref{on1}), and (\ref{on2}) are all ratios
of determinants, so that we can write 
\begin{equation}
\chi_{(n_{1}, n_{2}, \ldots, n_{N})}(M) = \frac{{\cal N}}{{\cal D}},
\end{equation}
where $M$ is a matrix element of the group in question and ${\cal N}$
and ${\cal D}$ refer to numerator and denominator.  For any of these
groups, the denominator ${\cal D}$ can be expressed generally as
\begin{equation}
{\cal D} = \det\left[t_{i}^{N-j+q}\pm t_{i}^{-(N-j+q)}-
\delta_{q0}\delta_{jN}\right],      \label{d}
\end{equation}
where $q$ can take on the value $1$ for the $Sp(2N)$ group,
$\frac{1}{2}$ for the $SO(2N+1)$ group and $0$ for the $SO(2N)$
group. In this form, we choose the appropriate value of $q$ and the
proper sign of the $\pm$ sign to specify which group we are
discussing.  Namely, we see that the minus sign will be used for
$Sp(2N)$ and $SO(2N+1)$ whereas the plus sign will be used for
$SO(2N)$. 

In following the derivation of the $U(N)$ expansion given in
Ref. \cite{Balantekin:2000vn}, we define a ``generating function'',
$G(x,t)$, to be some function of a variable $t$ and any necessary
parameters $x$.  Later, we will take $t$ to be an eigenvalue of a
group matrix.  For now, we expand the generating function in a power
series in the variable $t$ around $t=0$.  Thus, 
\begin{equation}
G(x,t)=\sum_{n=-{\infty}}^{\infty}A_{n}(x)t^{n}.   \label{g}
\end{equation}
We assume that the series expansion converges for $|t|=1$.  However,
there are no other restrictions on the coefficients, so that some of
the $A_{n}(x)$, can be zero.  For instance, if the expansion is a
Taylor Series, then $A_{n} \equiv 0$ for all $n < 0$.

Now, we define a function ${\cal F}$ by
\begin{equation}
{\cal F} = {\cal D}\prod_{i=1}^{N}G(x,t_{i})G(x,t_{i}^{-1}), \label{f}
\end{equation}
where $\cal D$ is given in Eq. (\ref{d}).  By using the definition  of
$G(x,t)$ from Eq. (\ref{g}), ${\cal F}$ can be written as (suppressing
the $x$ dependence of $A_{n}$)
\begin{eqnarray}
& {\cal F} =  {\cal D}
\left[{\displaystyle\sum_{n=-{\infty}}^{\infty}}A_{n}t_{1}^{n}\right]
\left[{\displaystyle\sum_{n=-{\infty}}^{\infty}}A_{n}t_{1}^{-n}\right]
\nonumber \\ &
\left[{\displaystyle\sum_{n=-{\infty}}^{\infty}}A_{n}t_{2}^{n}\right]
\left[{\displaystyle\sum_{n=-{\infty}}^{\infty}}A_{n}t_{2}^{-n}\right]
\cdots
\left[{\displaystyle\sum_{n=-{\infty}}^{\infty}}A_{n}t_{N}^{n}\right]
\left[{\displaystyle\sum_{n=-{\infty}}^{\infty}}A_{n}t_{N}^{-n}\right],
\end{eqnarray}
or by combining the product of sums over the same variable, it can be
written as
\begin{equation}
{\cal F} = {\cal D}\left[\sum_{n=-{\infty}}^{\infty}
\sum_{p=-{\infty}}^{\infty}A_{n}A_{p}t_{1}^{n-p}\right] \left
[\sum_{n=-{\infty}}^{\infty}\sum_{p=-{\infty}}^
{\infty}A_{n}A_{p}t_{2}^{n-p}\right] \cdots \left[
\sum_{n=-{\infty}}^{\infty}\sum_{p=-{\infty}}^{\infty}
A_{n}A_{p}t_{N}^{n-p}\right].   \label{mess}
\end{equation}

To proceed, we use the expression for ${\cal D}$ in Eq. (\ref{d}).
This determinant can be laboriously expanded as an alternating sum of
products of the elements (see Eq. (\ref{det})).  Upon doing so, we can
combine the factors of the variable $t_{i}$ in the determinant with
the factor in Eq. (\ref{mess}) of the same variable.  However,  before
naively doing so, we notice that there is a symmetry in the exponents
in the determinant.  We also notice that the double summations of the
$t_{i}$'s are unaffected by interchange of the dummy indices $n$ and
$p$.  So the symmetry of the exponents will be preserved if we use
$n-p$ in the product of the first term, $p-n$ in the product of the
second term, and split the delta term in half using $n-p$ in the first
one and $p-n$ in the second one. Upon doing so, we find that the new
expression is again a determinant. Rewriting this as a determinant, we
obtain 
\begin{equation}
{\cal F} = \det \left[ \sum_{n=-{\infty}}^{\infty}\sum_{p=-{\infty}}
^{\infty} A_{n}A_{p} \left(t_{i}^{N-j+q+n-p} \pm
t_{i}^{-(N-j+q+n-p)}-\frac{1}{2}\delta_{q0}\delta_{jN}(t_{i}^{n-p}+
t_{i}^{p-n})\right)\right].
\end{equation}
Now, we change variables, defining a new integer $r=N-j+n-p$.  This
gives 
\begin{equation}
{\cal F} = \det \left[ \sum_{p=-{\infty}}^{\infty}\sum_{r=-{\infty}}
^{\infty} A_{r-N+j+p}A_{p}\left(t_{i}^{r+q} \pm t_{i}^{-(r+q)}
-\frac{1}{2} \delta_{q0}\delta_{jN}(t_{i}^{r-N+j}+
t_{i}^{N-j-r})\right)\right].
\end{equation}
The order of the summation for $r$ and $p$ is interchangeable.  Also,
the delta term chooses only $N=j$.  Thus,
\begin{equation}
{\cal F} = \det \left[ \sum_{r=-{\infty}}^{\infty}\sum_{p=-{\infty}}
^{\infty} A_{r-N+j+p}A_{p}\left(t_{i}^{r+q} \pm t_{i}^{-(r+q)} -
\frac{1}{2}\delta_{q0}\delta_{jN}(t_{i}^{r}+t_{i}^{-r})\right)\right].
\end{equation}

We notice that all of the dependence on the dummy variable $p$ can be
isolated by defining
\begin{equation}
c_{r,j}=\sum_{p=-{\infty}}^{\infty}A_{r-N+j+p}A_{p}.   \label{ceqn}
\end{equation}
Also, by combining the delta term with the other term (again
remembering that it is only present for the $SO(2N)$ case in which
$q=0$ and we use the $+$ sign), we can write our expression for ${\cal
F}$ as 
\begin{equation}
{\cal F} = \det \left[ \sum_{r=-{\infty}}^{\infty}c_{r,j}
\left(t_{i}^{r+q} \pm t_{i}^{-(r+q)}\right)
\left(1-\frac{1}{2}\delta_{q0}\delta_{jN}\right)\right].
\label{feqn}
\end{equation}
We have come to the point in the derivation where it will be necessary
to specialize Eq. (\ref{feqn}) for the three different types of groups
by making appropriate choices for $q$ and the $\pm$ sign.  This will
be taken up in the next three sections.

\section{The Expansion over Sp(2N) Characters}
\label{sec-sp2n}

We begin with the $Sp(2N)$ case, as it is the simplest one.  The
starting point will be Eq. (\ref{feqn}).  For the $Sp(2N)$ case, $q=1$
and we choose the minus sign.  Thus, we have
\begin{equation}
{\cal F} = \det \left[ \sum_{r=-{\infty}}^{\infty}c_{r,j}
\left(t_{i}^{r+1} - t_{i}^{-(r+1)}\right)\right]
\end{equation}
where $c_{r,j}$ is defined in Eq. (\ref{ceqn}). Before proceeding,  it
is beneficial to change dummy indices again by letting $r+1
\rightarrow r$.  This gives us
\begin{equation}
{\cal F} = \det \left[ \sum_{r=-{\infty}}^{\infty}c_{r,j}^{\prime}
\left(t_{i}^{r} - t_{i}^{-r}\right)\right]
\end{equation}
where
\begin{equation}
c_{r,j}^{\prime}=\sum_{p=-{\infty}}^{\infty}A_{r-N+j-1+p}A_{p}.
\label{spc}
\end{equation}
This sum over $r$ from $-\infty$ to $\infty$ can be broken up into
positive $r$, negative $r$, and $r=0$. The $r=0$ term vanishes because
$t_{i}^{0}-t_{i}^{0}=0$. Then changing the negative values to positive
by replacing $r$ with $-r$ and collecting terms, we get 
\begin{equation}
{\cal F} = \det \left[ \sum_{r=0}^{\infty}d_{r,j}
\left(t_{i}^{r}-t_{i}^{-r}\right) \right],    \label{det1}
\end{equation}
where
\begin{equation}
d_{r,j} = c_{r,j}^{\prime}-c_{-r,j}^{\prime}.             \label{spd} 
\end{equation}
Eq. (\ref{det1}) is very similar to an expression that is treated in
Theorem 1.2.1 from Hua \cite{hua}.  We will need a slightly more
general form of this theorem, which we present in the appendix. Using
the result, Eq. (\ref{theorem}), we get
\begin{equation}
{\cal F} = \sum_{r_{1}>r_{2}>\cdots>r_{N} \geq 0}^{\infty}\det
\left[d_{r_{j},i}\right]\det\left[t_{i}^{r_{j}}-t_{i}^{-r_{j}}\right].
\end{equation}
Now, if in the summation, $r_{N}=0$, then both determinants vanish, so
we can restrict $r_{N} \geq 1$.  Let us define
\begin{equation}
r_{j}=n_{j}+N-j+1.
\end{equation}
Then, $r_{j} > r_{j+1}$ implies that $n_{j} \geq n_{j+1}$.
Furthermore, since $r_{N} \geq 1$, then $n_{N} \geq 0$.  Thus, we can
write the summation as
\begin{equation}
{\cal F} = \sum_{n_{1} \geq n_{2} \geq \cdots \geq n_{N} \geq 0}
^{\infty}\det\left[d_{n_{j}+N-j+1,i}\right]
\det\left[t_{i}^{n_{j}+N-j+1}-t_{i}^{-(n_{j}+N-j+1)}\right].
\end{equation}

In the above expression, the second determinant is seen to be exactly
the numerator in the Weyl formula for the characters of the symplectic
group given in Eq. (\ref{sp}).  We even have the appropriate
restrictions on the values of the $n_{i}$ that are necessary to make
the equation valid.  Thus, we can write the above expression as  
\begin{equation}
{\cal F} = \sum_{n_{1} \geq n_{2} \geq \cdots \geq n_{N} \geq 0}
^{\infty}\det\left[d_{n_{j}+N-j+1,i}\right]{\cal N}.
\end{equation}
Now we recall our definition of ${\cal F}$ from Eq. (\ref{f}).  If we
divide both sides by the denominator ${\cal D}$ and recall that our
expression for the character of the $Sp(2N)$ group is $\chi_{(n_{1},
n_{2}, \ldots, n_{N})}(P) =\frac{{\cal N}}{{\cal D}}$, then we obtain 
\begin{equation}
\prod_{i=1}^{N}G(x,t_{i})G(x,t_{i}^{-1})=\sum_{n_{1} \geq n_{2}  \geq
\cdots \geq n_{N} \geq 0}^{\infty}\det\left[d_{n_{j}+N-j+1,i}
\right]\chi_{(n_{1}, n_{2}, \ldots, n_{N})}(P).        \label{sp2nexp}
\end{equation}
This is our desired character expansion over the $Sp(2N)$ group!  It
is a sum over all irreducible representations of the symplectic
group. Expressions for the coefficients are obtained using Eqs.
(\ref{spd}) and (\ref{spc}). The result is 
\begin{equation}
d_{n_{j}+N-j+1,i} = \sum_{p=-{\infty}}^{\infty}A_{p}
\left(A_{n_{j}+i-j+p}-A_{-n_{j}-2N-2+i+j+p}\right).  \label{sp2ngen}
\end{equation}
In the special case in which the expansion of the generating function
$G(x,t)$ is a Taylor Series expansion with $A_{p} \equiv 0$ for all
$p<0$, this simplifies slightly to 
\begin{equation}
d_{n_{j}+N-j+1,i} = \sum_{p=0}^{\infty}A_{p}
\left(A_{p+|n_{j}+i-j|}-A_{p+n_{j}+2N+2-i-j}\right). \label{sp2ncoef}  
\end{equation}
We defer examples of the usage of this expansion until Section
\ref{sec-example}.  

\section{The Expansion Over SO(2N+1) Characters}
\label{sec-so2n1}

Once again, we start from Eq. (\ref{feqn}).  For $SO(2N+1)$, we have
$q=\frac{1}{2}$ and we choose the minus sign.  Then, we have
\begin{equation}
{\cal F} = \det \left[ \sum_{r=-{\infty}}^{\infty}c_{r,j}
\left(t_{i}^{r+\frac{1}{2}} - t_{i}^{-(r+\frac{1}{2})}\right)\right]
\end{equation}
and $c_{r,j}$ is defined in Eq. (\ref{ceqn}).

This sum over $r$ from $-\infty$ to $\infty$ can be broken up into
ranges of $r \geq 0$ and $r<0$, which gives
\begin{equation}
{\cal F} = \det \left[ \sum_{r=0}^{\infty}c_{r,j}\left(t_{i}^
{r+\frac{1}{2}}-t_{i}^{-(r+\frac{1}{2})}\right) + \sum_{r=-\infty}^
{-1}c_{r,j}\left(t_{i}^{r+\frac{1}{2}} - t_{i}^{-(r+\frac{1}{2})}
\right) \right].
\end{equation}
Changing variables in the second summation using $r \rightarrow
-(r+1)$ and collecting terms, we get 
\begin{equation}
{\cal F} = \det \left[ \sum_{r=0}^{\infty}d_{r,j}
\left(t_{i}^{r+\frac{1}{2}}-t_{i}^{-(r+\frac{1}{2})}\right) \right], 
\label{det2}
\end{equation}
where
\begin{equation}
d_{r,j} = c_{r,j}-c_{-r-1,j}.         \label{so1d}
\end{equation}
Once again, we refer to Eq. (\ref{theorem}) in the Appendix to
simplify Eq. (\ref{det2}) and we write 
\begin{equation}
{\cal F} = \sum_{r_{1}>r_{2}>\cdots>r_{N} \geq 0}^{\infty}\det
\left[d_{r_{j},i}\right]\det\left[t_{i}^{r_{j}+\frac{1}{2}}-
t_{i}^{-(r_{j}+\frac{1}{2})}\right].
\end{equation}
If we define
\begin{equation}
r_{j}=n_{j}+N-j,
\end{equation}
then the summation becomes
\begin{equation}
{\cal F} = \sum_{n_{1} \geq n_{2} \geq \cdots \geq n_{N} \geq 0}
^{\infty}\det\left[d_{n_{j}+N-j,i}\right]\det\left[t_{i}^
{n_{j}+N-j+\frac{1}{2}}-t_{i}^{-(n_{j}+N-j+\frac{1}{2})}\right].
\end{equation}
The second determinant is simply the numerator of the Weyl formula for
$SO(2N+1)$ as given in equation \ref{on1}, so using the definition of
$\cal F$  from Eq. (\ref{f}) and dividing by $\cal D$ gives  
\begin{equation}
\prod_{i=1}^{N}G(x,t_{i})G(x,t_{i}^{-1})=\sum_{n_{1} \geq n_{2}  \geq
\cdots \geq n_{N} \geq 0}^{\infty}\det\left[d_{n_{j}+N-j,i}
\right]\chi_{(n_{1}, n_{2}, \ldots, n_{N})}(R).       \label{son1exp}
\end{equation}
This is the expansion for the $SO(2N+1)$ group, again a sum over all
irreducible representations.  Note, however, that this expression does
not include the spinor representations of $SO(2N+1)$.  The expression
for the coefficient is found using Eqs. (\ref{so1d}) and (\ref{ceqn})
with the result given by 
\begin{equation}
d_{n_{j}+N-j,i} = \sum_{p=-{\infty}}^{\infty}A_{p}
\left(A_{n_{j}+i-j+p}-A_{-n_{j}-2N-1+i+j+p}\right).   \label{so1gen}
\end{equation}
In the special case of a Taylor Series with $A_{p} \equiv 0$ for all
$p<0$, this simplifies to 
\begin{equation}
d_{n_{j}+N-j,i} = \sum_{p=0}^{\infty}A_{p}
\left(A_{p+|n_{j}+i-j|}-A_{p+n_{j}+2N+1-i-j}\right).   \label{so1coef}
\end{equation}

We conclude this section with a reminder that care must be taken in
the usage of the above formulas for $SO(2N+1)$.  One must remember
that the number 1 is always an additional eigenvalue of the matrix
$R$.  Thus, in forming group functions, one must manually include a
factor of $G(x,1)$ on both sides of the equation in order to have a
function on the left hand side that treats all eigenvalues equally.
This tricky point will be illustrated by example in Section
\ref{sec-example}, after we treat the $SO(2N)$ case in the next
section. 

\section{The Expansion Over SO(2N) Characters}
\label{sec-so2n}

One more time, we start from equation \ref{feqn}.  Recall that for
$SO(2N)$, we have $q=0$ and we use the $+$ sign.  Thus, we have
\begin{equation}
{\cal F} = \det \left[ \sum_{r=-{\infty}}^{\infty}c_{r,j}
\left(t_{i}^{r} + t_{i}^{-r}\right)\left(1-\frac{1}{2}
\delta_{jN}\right)\right]
\end{equation}
where $c_{r,j}$ is defined in equation \ref{ceqn}. The delta function
term serves to divide each entry in the last column by a factor of
two.  When taking the determinant, a factor of two comes out and
divides the equation.  Thus, 
\begin{equation}
{\cal F} = \frac{1}{2} \det \left[ \sum_{r=-\infty}^{\infty}c_{r,j}
\left(t_{i}^{r}+t_{i}^{-r}\right) \right].
\end{equation}

This sum over $r$ from $-\infty$ to $\infty$ can be broken up into
positive $r$, negative $r$, and $r=0$.  Then changing the negative
values to positive by replacing $r$ with $-r$ and collecting terms, 
we get 
\begin{equation}
{\cal F} = \frac{1}{2}\det \left[ \sum_{r=0}^{\infty}d_{r,j}
\left(1-\frac{1}{2}\delta_{r0}\right)
\left(t_{i}^{r}+t_{i}^{-r}\right) \right],      \label{det3}
\end{equation}
where
\begin{equation}
d_{r,j} = c_{r,j}+c_{-r,j}         \label{so2d}
\end{equation}
and the $\delta_{r0}$ is inserted to ensure the correct coefficient
for $r=0$.  Once again, we can use Eq. (\ref{theorem}) from the
appendix to simplify Eq. (\ref{det3}) which gives 
\begin{equation}
{\cal F} = \frac{1}{2} \sum_{r_{1}>r_{2}>\cdots>r_{N} 
\geq 0}^{\infty}
\det\left[d_{r_{j},i}\right]
\det\left[\left(t_{i}^{r_{j}}+t_{i}^{-r_{j}}\right)
\left(1-\frac{1}{2} 
\delta_{r_{j}0}\right)\right].
\end{equation}

Let us define
\begin{equation}
r_{j}=n_{j}+N-j.
\end{equation}
Then the summation becomes
\begin{equation}
{\cal F} = \frac{1}{2}\sum_{n_{1} \geq n_{2} \geq \cdots \geq n_{N}
\geq 0}^{\infty} \det\left[d_{n_{j}+N-j,i}\right]\det\left[\left(
t_{i}^{n_{j}+N-j}+t_{i}^{-(n_{j}+N-j)}\right)\left(1-\frac{1}{2}
\delta_{n_{j}+N-j,0}\right)\right]. \label{function}
\end{equation}
Focusing on the second determinant on the right hand side, we can
multiply the two binomials to give
\begin{equation}
\det\left[t_{i}^{n_{j}+N-j}+t_{i}^{-(n_{j}+N-j)}-\frac{1}{2}
\delta_{n_{j}+N-j,0}\left(t_{i}^{n_{j}+N-j}+t_{i}^{-(n_{j}+N-j)}
\right)\right]. 
\end{equation}
Now, the delta function is only non-zero when $n_{j}+N-j=0$, which can
only occur for $j=N$ and $n_{N}=0$ because $n_{j}$ is non-negative.
In this event, the exponents vanish and the sum in parentheses becomes
2, which cancels the $\frac{1}{2}$.  Thus, we can write Eq. 
(\ref{function}) as 
\begin{equation}
{\cal F} = \frac{1}{2}\sum_{n_{1} \geq n_{2} \geq \cdots \geq n_{N}
\geq 0}^{\infty} \det\left[d_{n_{j}+N-j,i}\right] \det\left[
t_{i}^{n_{j}+N-j}+t_{i}^{-(n_{j}+N-j)}-\delta_{jN}\delta_{n_{N}0}
\right].
\end{equation}

We see that the second determinant is precisely the appropriate
numerator in the Weyl formula for the quantity  ${\cal C}_{(n_{1},
n_{2}, \ldots,  n_{N})}(R)$ given in Eq. (\ref{on2}).  By recalling
the definition of $\cal F$ from Eq. (\ref{f}) and dividing by $\cal
D$, we get 
\begin{equation}
\prod_{i=1}^{N}G(x,t_{i})G(x,t_{i}^{-1})=\sum_{n_{1}\geq n_{2}\geq
\cdots \geq n_{N}\geq 0}^{\infty} \frac{1}{2}
\det\left[d_{n_{j}+N-j,i} \right]{\cal C}_{(n_{1}, n_{2}, \ldots,
n_{N})}(R),    \label{so2nexp}
\end{equation}
This is the character expansion for $SO(2N)$.  As with the $SO(2N+1)$  
case, the expansion does not include the spinor representations.
Note that the above expansion is not an expansion
over the simple characters of the $SO(2N)$ group because the ${\cal
C}$'s are double characters if $n_{N}>0$ as discussed in Section
\ref{sec-back}.  If one desires an expansion over the simple
characters, one can write 
\begin{equation}
{\cal C} = \frac{1}{2}\left({\cal C} + {\cal S}\right) +
\frac{1}{2}\left({\cal C} - {\cal S}\right),
\end{equation}
which puts the two simple characters on the right hand side, as
explained earlier.  At the present time, we find it simpler to apply
the formula in the state that it is in.  The expression for the
coefficient is found using Eqs. (\ref{so2d}) and (\ref{ceqn}) and is
found to be 
\begin{equation}
d_{n_{j}+N-j,i} = \sum_{p=-{\infty}}^{\infty}A_{p}
\left(A_{n_{j}+i-j+p}+A_{-n_{j}-2N+i+j+p}\right).    \label{so2gen}
\end{equation}
In the special case of a Taylor Series with $A_{p}=0$ for all $p<0$,
this simplifies to 
\begin{equation}
d_{n_{j}+N-j,i} = \sum_{p=0}^{\infty}A_{p}\left(A_{p+|n_{j}+i-j|}+
A_{p+n_{j}+2N-i-j}\right).             \label{so2coef}
\end{equation}
The derivations of the expansions are complete.  We now turn to some
examples.

\section{Examples of Character Expansions}
\label{sec-example}

In this section, we give some examples of expansions of group
functions over the characters of the $Sp(2N)$, $SO(2N+1)$, and
$SO(2N)$ groups.  In subsection A, we will present the expansion of
the generating function of the homogeneous symmetric functions.  This
can be used as a check of the formulas derived in the present paper,
as the expansions are known.  In subsection B, we present the
expansion for the function $\exp(x Tr M)$, where $M$ is some matrix
element of one of the three groups we treat. 

\subsection{Homogeneous Symmetric Functions}

Consider the generating function of the homogeneous symmetric
functions, namely
\begin{equation}
G(x,t)=\frac{1}{1-xt}=\sum_{n=0}^{\infty}x^{n}t^{n}.  \label{hgen}
\end{equation}
Thus, we have $A_{n}(x)=x^{n}$ for $n \geq 0$ and $A_{n}(x)=0$
otherwise.  Consider the $Sp(2N)$ expansion.  Note that 
\begin{equation}
\prod_{i=1}^{N}G(x,t_{i})G(x,t_{i}^{-1}) 
=\frac{1}{\det\left[I-xP\right]}
\end{equation}
where $I$ is the $2N \times 2N$ identity matrix.  The expansion is
given by Eq. (\ref{sp2nexp}).  Since the series expansion of the
generating function in Eq. (\ref{hgen}) does not contain negative
powers of $t$, the coefficients are given by Eq. (\ref{sp2ncoef}).
Thus, the coefficients are given by
\begin{equation}
\det\left[d_{n_{j}+N-j+1,i}\right] =
\det\left[\sum_{p=0}^{\infty}x^{p}
\left(x^{p+|n_{j}+i-j|}-x^{p+n_{j}+2N+2-i-j}\right)\right],
\end{equation}
which after simplifying becomes 
\begin{equation}
\det\left[d_{n_{j}+N-j+1,i}\right] = \det\left[\frac{x^{|n_{j}+i-j|}-
x^{n_{j}+2N+2-i-j}}{1-x^{2}}\right].
\end{equation}
We simplify by noticing that if $n_{2} \geq 1$, then the first column
of the determinant is a multiple of the second column,
thereby making the determinant vanish.  Thus, $n_{2}$ must be zero in
order to have a non-vanishing coefficient.  Now, since $n_{2} \geq
n_{3}$ and so on, we see that the only surviving terms in the
expansion are those for which $n_{2}=n_{3}=\cdots = n_{N}=0$.  This
corresponds to one row Young tableaux, labeled $(n)$ earlier.  The
above determinant then becomes
\begin{equation}
\det\left[d_{n_{j}+N-j+1,i}\right] =
\det\left[\frac{x^{|n_{1}\delta_{1j}
+i-j|}-x^{n_{1}\delta_{1j}+2N+2-i-j}}{1-x^{2}}\right].
\end{equation}
This, in turn, can be written as
\begin{equation}
\det\left[d_{n_{j}+N-j+1,i}\right] = \frac{x^{n_{1}}}{(1-x^{2})^{N}}
\det\left[x^{|i-j|}-x^{2N+2-i-j}\right].
\end{equation}
By an induction argument on the dimension of the determinant on the
right hand side, one can prove that
\begin{equation}
\det\left[x^{|i-j|}-x^{2N+2-i-j}\right]=(1-x^{2})^{N}
\end{equation}
and thus
\begin{equation}
\det\left[d_{n_{j}+N-j+1,i}\right] = x^{n_{1}},
\end{equation}
where we recall that all $n$'s other than $n_{1}$ are 0.  Then,
the character expansion is Eq. (\ref{sp2nexp}) with the coefficients
found above is 
\begin{equation}
\frac{1}{\det\left[I-xP\right]}=\prod_{i=1}^{N}
\left(\frac{1}{1-xt_{i}}\right)\left(\frac{1}
{1-xt_{i}^{-1}}\right)=\sum_{n_{1}=0}^{\infty}x^{n_{1}}
\chi_{(n_{1})}(P).
\end{equation}
If we use Eq. (\ref{hasp}) which relates the character of one row
Young Tableaux to the homogeneous symmetric functions, we have
\begin{equation}
\frac{1}{\det\left[I-xP\right]}=\prod_{i=1}^{N}
\left(\frac{1}{1-xt_{i}}\right)\left(\frac{1}
{1-xt_{i}^{-1}}\right)=\sum_{n=0}^{\infty}x^{n}h_{n}(t_{i},
t_{i}^{-1}).
\end{equation}
However, this is exactly the defining equation for the homogeneous
symmetric functions.  Thus, we see that the expansion derived for
$Sp(2N)$ agrees with the known expansion. 

To perform the same expansion over the $SO(2N+1)$ group, one must use 
caution.  As alluded to earlier, we must manually include the
eigenvalue 1.  Mathematically, we have
\begin{equation}
\frac{1}{\det\left[I-xR\right]}=G(x,1)\prod_{i=1}^{N}
G(x,t_{i})G(x,t_{i}^{-1}),
\end{equation}
where $G(x,t)$ is still given by Eq. (\ref{hgen}) and $I$ is the
$(2N+1) \times (2N+1)$ identity matrix.  Then, we have the
expansion from Eq. (\ref{son1exp}) and we scale both sides by
$G(x,1)=(1-x)^{-1}$ to get
\begin{equation}
G(x,1)\prod_{i=1}^{N}G(x,t_{i})G(x,t_{i}^{-1})=\frac{1}{1-x}
\sum_{n_{1} \geq n_{2} \geq \cdots \geq n_{N} \geq 0}^{\infty}
\det\left[d_{n_{j}+N-j,i}\right]\chi_{(n_{1}, n_{2}, \ldots,
n_{N})}(R).
\end{equation}
The coefficient is given by Eq. (\ref{so1coef}) 
\begin{equation}
\det\left[d_{n_{j}+N-j,i}\right] = \det\left[\sum_{p=0}^{\infty}x^{p}
\left(x^{p+|n_{j}+i-j|}-x^{p+n_{j}+2N+1-i-j}\right)\right]
\end{equation}
or after simplifying,
\begin{equation}
\det\left[d_{n_{j}+N-j,i}\right] = \det\left[\frac{x^{|n_{j}+i-j|}
-x^{n_{j}+2N+1-i-j}}{1-x^{2}}\right].
\end{equation}
Once again it can be shown that if $n_{2} \geq 1$, the first two
columns are multiples and the coefficient vanishes.  It can also be
shown using the result of the similar expression for the $Sp(2N)$
expansion that if $n_{1}=n$ and all other $n_{i}=0$, then
\begin{equation}
\det\left[d_{n_{j}+N-j,i}\right] = \frac{x^{n}}{1+x}
\end{equation}
so that 
\begin{equation}
\frac{1}{\det\left[I-xR\right]}=\frac{1}{\left(1-x\right)}
\prod_{i=1}^{N}\frac{1}{\left(1-xt_{i}\right)}\frac{1}
{\left(1-xt_{i}^{-1}\right)}=\sum_{n=0}^{\infty}\frac{x^{n}}{1-x^{2}}
\chi_{(n)}(R).
\end{equation}
Finally, using the value of the single row characters for the
$SO(2N+1)$ group from Eq. (\ref{haso2n1}), one can show that
\begin{equation}
\frac{1}{\det\left[I-xR\right]}=\frac{1}{\left(1-x\right)}
\prod_{i=1}^{N}\frac{1}{\left(1-xt_{i}\right)}\frac{1}
{\left(1-xt_{i}^{-1}\right)}=\sum_{n=0}^{\infty}x^{n}
h_{n}(t_{i},t_{i}^{-1},1),
\end{equation}
which again agrees with the definition of the homogeneous symmetric
functions.

We continue on with the same expansion for the $SO(2N)$ group. Using
the same generating function and using the character expansion from
Eq. (\ref{so2nexp}), we have
\begin{eqnarray}
{\displaystyle\frac{1}{\det[I-xR]}} &=&{\displaystyle\prod_{i=1}^{N}
\frac{1}{\left(1-xt_{i}\right)}\frac{1}{\left(1-xt_{i}^{-1}\right)}}
\nonumber \\ &=& {\displaystyle\sum_{n_{1} \geq n_{2} \geq \cdots \geq
n_{N} \geq 0} ^{\infty}\frac{1}{2}}\det\left[d_{n_{j}+N-j,i}\right]
{\cal C}_{(n_{1}, n_{2}, \ldots, n_{N})}(R).
\end{eqnarray}
Note the appearance of the factor of $\frac{1}{2}$ in this expression.
The coefficient is given by Eq. (\ref{so2coef}) to be 
\begin{equation}
\det\left[d_{n_{j}+N-j,i}\right] = \det\left[\sum_{p=0}^{\infty}x^{p}
\left(x^{p+|n_{j}+i-j|}+x^{p+n_{j}+2N-i-j}\right)\right],             
\end{equation}
or after simplification,
\begin{equation}
\det\left[d_{n_{j}+N-j,i}\right] = \det\left[\frac{x^{|n_{j}+i-j|}+
x^{n_{j}+2N-i-j}}{1-x^{2}}\right].
\end{equation}
Once again, the first two columns are multiples if $n_{2} \geq 1$, so
the only surviving coefficients are the ones corresponding to
$n_{1}=n$, all others are zero.  Again, the determinant can be
evaluated with the help of the result from the $Sp(2N)$ case, and the
result is 
\begin{equation}
\det\left[d_{n_{j}+N-j,i}\right] = \frac{2x^{n}}{1-x^{2}}.
\end{equation}
We note that the 2 cancels with the $\frac{1}{2}$ built in to the
$SO(2N)$ expansion and the remaining expression is exactly the same
as the $SO(2N+1)$ expression.  Thus, we see that the expansions
derived in the present work indeed give the correct expansion.  In all
three of these examples, we have tacitly assumed that $N$ is at least
2, but one can check that the expansions are correct for the $N=1$
cases as well.

We could also consider the generating function for the alternating
symmetric functions, $G(x,t)=1-xt$, and calculate the expansions as
another check for reliability.  One can check that the expansions
derived from the present work in fact give the known expansions.  This
task will not be undertaken in the present work.

\subsection{Expansion of exp(x Tr M)}

Now that we have confidence in the character expansions derived here,
we can start considering more interesting examples.  Of course, any
generating function can be chosen as long as it can be expanded in a
power series.  For our example, we will choose the exponential
function because it is expected that this technique will prove useful
in performing group integrals that arise in low-energy effective QCD
partition functions and the integrals are of exponential functions.

We begin by defining the generating function
\begin{equation}
G(x,t)=e^{xt}=\sum_{n=0}^{\infty}\frac{x^{n}}{n!}t^{n}
\end{equation}
so that $A_{n}(x)=\frac{x^{n}}{n!}$ for $n \geq 0$ and zero
otherwise. Let us first consider $Sp(2N)$.  We have 
\begin{equation}
\prod_{i=1}^{N}G(x,t_{i})G(x,t_{i}^{-1})=e^{x\left(t_{1}+t_{2}
+\cdots+t_{N}+t_{1}^{-1}+t_{2}^{-1}+\cdots+t_{N}^{-1}\right)}= \exp (x
Tr P).
\end{equation}
Our character expansion is given by Eq. (\ref{sp2nexp}) as
\begin{equation}
\exp (x Tr P)=\sum_{n_{1} \geq n_{2} \geq \cdots \geq n_{N} \geq 0}
^{\infty}\det\left[d_{n_{j}+N-j+1,i}\right] \chi_{(n_{1}, n_{2},
\ldots, n_{N})}(P) , 
\end{equation}
where the coefficients are given directly by Eq. (\ref{sp2ncoef}) as 
\begin{equation}
\det\left[d_{n_{j}+N-j+1,i}\right] = \det\left[\sum_{p=0}^{\infty}
\frac{x^{p}}{p!}\left(\frac{x^{p+|n_{j}+i-j|}}{\left(p+|n_{j}+i+j|
\right)!}-\frac{x^{p+n_{j}+2N+2-i-j}}{\left(p+n_{j}+2N+2-i-j\right)!}
\right)\right].
\end{equation}
This can be recognized as a modified Bessel function, which has the
expansion
\begin{equation}
I_{\lambda}(x)=\sum_{p=0}^{\infty}\frac{1}{p!(p+\lambda)!}
\left(\frac{x}{2}\right)^{2p+\lambda}.
\end{equation}
Note also that $I_{\lambda}(x)=I_{-\lambda}(x)$ for any $x$.  Thus, we
can rewrite the coefficient as (dropping the absolute value sign)
\begin{equation}
\det\left[d_{n_{j}+N-j+1,i}\right] = \det\left[I_{n_{j}+i-j}(2x)-
I_{n_{j}+2N+2-i-j}(2x)\right]
\end{equation}
so that finally,
\begin{equation}
\exp (x Tr P)=\sum_{n_{1} \geq n_{2} \geq \cdots \geq n_{N} \geq 0}
^{\infty}\det\left[I_{n_{j}+i-j}(2x)- I_{n_{j}+2N+2-i-j}(2x)\right]
\chi_{(n_{1}, n_{2}, \ldots, n_{N})}(P) . 
\end{equation}

We proceed with the same expansion for the $SO(2N+1)$ group.  Using
the same generating function, Eq. (\ref{son1exp}) gives us the
expansion as
\begin{equation}
\prod_{i=1}^{N} \exp(xt_{i}) \exp(xt_{i}^{-1})=\sum_{n_{1} \geq n_{2}
\geq  \cdots \geq n_{N} \geq
0}^{\infty}\det\left[d_{n_{j}+N-j,i}\right] \chi_{(n_{1}, n_{2},
\ldots, n_{N})}(R),
\end{equation}
where Eq. (\ref{so1coef}) gives
\begin{equation}
\det\left[d_{n_{j}+N-j,i}\right] = \det\left[\sum_{p=0}^{\infty}
\frac{x^{p}}{p!}\left(\frac{x^{p+|n_{j}+i-j|}}
{\left(p+|n_{j}+i-j|\right)!}-\frac{x^{p+n_{j}+2N+1-i-j}}
{\left(p+n_{j}+2N+1-i-j\right)!}\right)\right],
\end{equation}
or using the definition of the modified Bessel function, 
\begin{equation}
\det\left[d_{n_{j}+N-j,i}\right] = \det\left[I_{n_{j}+i-j}(2x)-
I_{n_{j}+2N+1-i-j}(2x)\right].
\end{equation}
Thus, our expression is
\begin{equation}
\prod_{i=1}^{N} \exp (xt_{i}) \exp (xt_{i}^{-1})=\sum_{n_{1} \geq
n_{2}  \geq \cdots \geq n_{N} \geq 0}^{\infty}
\det\left[I_{n_{j}+i-j}(2x)- I_{n_{j}+2N+1-i-j}(2x)\right]
\chi_{(n_{1}, n_{2}, \ldots, n_{N})}(R). 
\end{equation}
Now, the left hand side is not yet $\exp (xTrR)$.  We need to include
the eigenvalue 1.  Thus we multiply both sides by $e^{x}$ and we get
\begin{eqnarray}
&\exp (x Tr R)= \exp(x){\displaystyle\prod_{i=1}^{N}} \exp(xt_{i})
\exp(xt_{i}^{-1})= \nonumber \\ & e^{x}{\displaystyle\sum_{n_{1}\geq
n_{2}\geq \cdots\geq n_{N}\geq 0}
^{\infty}}\det\left[I_{n_{j}+i-j}(2x)-I_{n_{j}+2N+1-i-j}(2x)\right]
\chi_{(n_{1}, n_{2}, \ldots, n_{N})}(R),
\end{eqnarray}
which is the desired expansion.  We emphasize the appearance of the
$e^{x}$ on the right hand side of the expression. This extra term is
unique to the $SO(2N+1)$ group.

As our final example, we develop the same expansion for the $SO(2N)$
group.  As before, we write the expansion from Eq. (\ref{so2nexp})  as
\begin{equation}
\exp(x Tr R)=\prod_{i=1}^{N} \exp(xt_{i})
\exp(xt_{i}^{-1})=\sum_{n_{1}   \geq n_{2} \geq \cdots \geq n_{N} >
0}^{\infty} \frac{1}{2} \det \left[  d_{n_{j}+N-j,i}\right]{\cal
C}_{(n_{1}, n_{2}, \ldots, n_{N})}(R) . 
\end{equation}
The coefficients are given by Eq. (\ref{so2coef}) as
\begin{equation}
\det\left[d_{n_{j}+N-j,i}\right] =\det\left[\sum_{p=0}^{\infty}
\frac{x^{p}}{p!}\left(\frac{x^{p+|n_{j}+i-j|}}
{\left(p+|n_{j}+i-j|\right)!}+\frac{x^{p+n_{j}+2N-i-j}}
{\left(p+n_{j}+2N-i-j\right)!}\right)\right].
\end{equation}
Again using the modified Bessel equation expansion, we get
\begin{equation}
\det\left[d_{n_{j}+N-j,i}\right] = \det\left[I_{n_{j}+i-j}(2x)+
I_{n_{j}+2N-i-j}(2x)\right].
\end{equation}
Thus, the desired expansion is
\begin{equation}
\exp(x Tr R)=\sum_{n_{1} \geq n_{2} \geq \cdots \geq n_{N} \geq 0}
^{\infty}\frac{1}{2}\det\left[I_{n_{j}+i-j}(2x)+I_{n_{j}+2N-i-j}
(2x)\right]{\cal C}_{(n_{1}, n_{2}, \ldots, n_{N})}(R).
\end{equation}
Once again, we emphasize the factor of $\frac{1}{2}$ in this
expression.  This is unique to the $SO(2N)$ expansion.

\section{Conclusions}

The present paper, along with Refs. \cite{Balantekin:2000vn} and
\cite{Balantekin:1984km} completes the program of finding character
expansions for all classical Lie groups. We expect these formulas to
be useful in a wide range of applications. We already described some
of these applications in the Introduction. 

One should emphasize that the success in understanding the
relationship between the random matrix theories and the low-lying
eigenvalues of the QCD Dirac operator suggest investigating other
aspects of QCD in a statistical framework (for a recent review see
Ref. \cite{Verbaarschot:2000dy}). More recently a similarity between
disordered systems in condensed matter physics and QCD, namely the
existence of a universal energy scale known as Thouless energy, was
suggested \cite{Janik:1998ki,Osborn:1998nm,Guhr:2000cf}. This problem
can be treated using the supersymmetry approach
\cite{efetov,Guhr:1998ve,Verbaarschot:1985jn}. In the supersymmetry
approach to this problem one needs to calculate integrals over
supergroups \cite{Guhr:2001ze,guhr1991}. One should note that
integration over unitary supergroups was already considered in
Refs. \cite{Guhr:1996vx} and
\cite{guhr1991,Guhr:1993ei,Alfaro:1995ca}.  Invariant integration over
an $Osp(N/2M)$ manifold was also  previously discussed in
Refs. \cite{Balantekin:1988rp} and \cite{Balantekin:1989gw}.  An
approach based on Gelfand-Tzetlin coordinates was developed and a
recursion formula for both ordinary and supergroup integrals was found
\cite{guhr1996,kohler,k1,k2}. Character expansions for supergroups may
be useful to understand the nature and extent of this approach. The
characters of supergroups are given by formulas similar to the Weyl
formulas except that complete symmetric functions are replaced by the
graded homogeneous symmetric functions or alternately traces by
supertraces \cite{BahaBalantekin:1981qy,BahaBalantekin:1981pp,%
BahaBalantekin:1982ku,BahaBalantekin:1982bk}. Since our character
expansion formulas are basically combinatorial in nature they are
applicable to the supergroups as well by the appropriate substitution
of traces with supertraces. Thus one can obtain character expansions
of the orthosymplectic supergroup $Osp(N/2M)$ from our formulas for
$SO(N)$ \cite{BahaBalantekin:1981qy} and of the supergroup $P(N)$ from
our formulas for $Sp(2N)$ \cite{BahaBalantekin:1981pp}.

\section*{ACKNOWLEDGMENTS}

We thank Gernot Akemann for useful discussions. This work was
supported in part by the U.S. National  Science Foundation Grant No.\
PHY-0070161  at the University of Wisconsin, in part by the University
of Wisconsin Research Committee with funds granted by the Wisconsin
Alumni Research Foundation, and in part by the Alexander von
Humboldt-Stiftung.  A.B.B. is grateful to the Max-Planck-Institut
f\"ur Kernphysik and H.A. Weidenm\"uller for the very kind
hospitality.

\appendix
\section{A Theorem on Determinants}
\label{sec-det}

Here we take up the issue of slightly generalizing a theorem on
determinants that can be found in Hua's book \cite{hua}.  His Theorem
1.2.1 states that
\begin{equation}
\det\left[\sum_{r=0}^{\infty}d_{r,j}t_{i}^{r}\right]=
\sum_{r_{1}>r_{2}>\ldots>r_{N}\geq 0}^{\infty}
\det\left[d_{r_{j},i}\right]\det\left[t_{i}^{r_{j}}\right].
\end{equation}
We would like to prove the following more general statement, which we
state as a theorem.  \newtheorem{huas}{Theorem}
\begin{huas}
Let $f_{r}(t)$ be an arbitrary function of the variable $t$ with
dependence on the index $r$.  Then the following equality holds:  
\begin{equation}
\det\left[\sum_{r=0}^{\infty}d_{r,j}f_{r}(t_{i})\right]=
\sum_{r_{1}>r_{2}>\ldots>r_{N}\geq 0}^{\infty}
\det\left[d_{r_{j},i}\right]\det\left[f_{r_{j}}(t_{i})\right].  
\label{thm}
\end{equation}
\end{huas}

To prove the theorem, we recall the expansion of determinants of $N
\times N$ matrices, namely
\begin{equation}
\det\left[{\cal A}_{i,j}\right]=\sum_{m_{1}=1}^{N}\sum_{m_{2}=1}^{N}
\cdots\sum_{m_{N}=1}^{N}\epsilon_{m_{1} m_{2} \cdots m_{N}} {\cal
A}_{1,m_{1}}{\cal A}_{2,m_{2}}\cdots{\cal A}_{N,m_{N}} ,  \label{det}
\end{equation}
where the tensor $\epsilon_{m_{1} m_{2} \cdots m_{N}}$ is completely
antisymmetric.  Then, the left hand side of Eq. (\ref{thm}) becomes
\begin{eqnarray}
& \det\left[{\displaystyle\sum_{r=0}^{\infty}}d_{r,j}f_{r}
(t_{i})\right]={\displaystyle\sum_{m_{1}=1}^{N}\sum_{m_{2}=1}^{N}
\cdots\sum_{m_{N}=1}^{N}}\epsilon_{m_{1} m_{2} \cdots m_{N}}
\left[{\displaystyle\sum_{r_{1}=0}^{\infty}}d_{r_{1},m_{1}}
f_{r_{1}}(t_{1})\right] \nonumber \\ &
\left[{\displaystyle\sum_{r_{2}=0}^{\infty}}d_{r_{2},m_{2}}
f_{r_{2}}(t_{2})\right] \cdots
\left[{\displaystyle\sum_{r_{N}=0}^{\infty}}d_{r_{N},m_{N}}
f_{r_{N}}(t_{N})\right] .
\end{eqnarray}
Isolating the dependence on the $m_{i}$'s, we get
\begin{eqnarray}
& \det\left[{\displaystyle\sum_{r=0}^{\infty}}d_{r,j}f_{r}
(t_{i})\right]={\displaystyle\sum_{r_{1}=0}^{\infty}
\sum_{r_{2}=0}^{\infty}\cdots\sum_{r_{N}=0}^{\infty}}
f_{r_{1}}(t_{1})f_{r_{2}}(t_{2})\cdots f_{r_{N}}(t_{N}) \nonumber \\ &
\left[{\displaystyle\sum_{m_{1}=1}^{N}\sum_{m_{2}=1}^{N}
\cdots\sum_{m_{N}=1}^{N}}\epsilon_{m_{1} m_{2} \cdots m_{N}}
d_{r_{1},m_{1}}d_{r_{2},m_{2}}\cdots d_{r_{N},m_{N}} \right].
\end{eqnarray}
We recognize the term on the right hand side in the large brackets as
a determinant, so 
\begin{equation}
\det\left[\sum_{r=0}^{\infty}d_{r,j}f_{r}(t_{i})\right]=
\sum_{r_{1}=0}^{\infty}\sum_{r_{2}=0}^{\infty}\cdots
\sum_{r_{N}=0}^{\infty}f_{r_{1}}(t_{1})f_{r_{2}}(t_{2}) \cdots
f_{r_{N}}(t_{N})\det\left[d_{r_{i},j}\right].
\end{equation}

Now, if any of the $r_{i}$ are equal, then the determinant on the
right hand side will vanish because two rows would be identical.
Thus, the sum can be restricted to distinct values of the $r_{i}$'s.
Next, since the $r_{i}$'s are all different, we would like to order
them in descending order so that $r_{i} > r_{i+1}$.  In doing so, we
would like to not change the form of the determinant on the right hand
side.  So, for any switch of labels, we permute the rows to leave the
form unchanged.  This brings in a factor of $+1$ or $-1$, depending on
how many permutations are needed.  We can express this simply by using
the N-th rank alternating tensor as
\begin{eqnarray}
&\det\left[{\displaystyle\sum_{r=0}^{\infty}}d_{r,j}f_{r}(t_{i})\right]
={\displaystyle\sum_{r_{1} > r_{2} > \cdots > r_{N} \geq 0}^{\infty}}
\det\left[d_{r_{i},j}\right] \nonumber \\ &
{\displaystyle\sum_{m_{1}=1}^{N}\sum_{m_{2}=1}^{N}
\cdots\sum_{m_{N}=1}^{N}}\epsilon_{m_{1} m_{2} \cdots m_{N}}
f_{r_{m_{1}}}(t_{1})f_{r_{m_{2}}}(t_{2}) \cdots f_{r_{m_{N}}}(t_{N}).
\end{eqnarray}
Here we note that the second line is a determinant.  Thus, taking the
transpose of the determinant of the $d$'s, we conclude
\begin{equation}
\det\left[\sum_{r=0}^{\infty}d_{r,j}f_{r}(t_{i})\right]= \sum_{r_{1} >
r_{2} > \cdots > r_{N} \geq 0}^{\infty} \det\left[d_{r_{j},i}\right]
\det\left[f_{r_{j}}(t_{i})\right],
\label{theorem}
\end{equation}
which proves the theorem.

\end{document}